\documentclass{article}
\usepackage{fortschritte}
\usepackage{epsfig,amssymb,euscript}
\usepackage{amsmath,amscd}

\newcommand{\be}{\begin{eqnarray}}
\newcommand{\ee}{\end{eqnarray}}
\newcommand{\bea}{\begin{eqnarray}}
\newcommand{\eea}{\end{eqnarray}}
\newcommand{\ba}{\begin{array}}
\newcommand{\ea}{\end{array}}

\newcommand{\nn}{\nonumber \\}

\newcommand{\p}[1]{(\ref{#1})}

\font\mybb=msbm10 at 11pt 
\def\bb#1{\hbox{\mybb#1}}

\def\bR {\bb{R}}
\def\bM {\bb{M}}
\def\bC {\bb{C}}

\newcommand{\bomega}{{\mbox{\boldmath $\omega$}}}

\def\beq{\begin{equation}}                     %
\def\eeq{\end{equation}}                       %
\def\bea{\begin{eqnarray}}                     
\def\eea{\end{eqnarray}}                       
\begin {document}
\def\titleline{
%
%
%
Classifying Supergravity Solutions
}
\def\email_speaker{
{\tt
%
%
}}
\def\authors{
%
%
%
%
%
Jerome P. Gauntlett
}
\def\addresses{
%
%
%
%
Blackett Laboratory, Imperial College \\            %
Prince Consort Rd., London, SW7 2AZ, U.K.\\            
                                            %
}
\def\abstracttext{
%
%
We review the substantial progress that has been made in
classifying supersymmetric solutions of supergravity theories
using $G$-structures. We also review the construction of
supersymmetric black rings that were discovered using the
classification of $D=5$ supergravity solutions.

{\it To appear in the Proceedings of the 37th International
Symposium Ahrenshoop, ``Recent Developments in String/M-Theory and
Field Theory", August 23-27 2004, Berlin-Schm\"ockwitz.}

}
\large
\makefront
\section{Introduction}

Supersymmetric solutions of supergravity theories have played a
key role in many of the most important developments in string
theory. For example, supersymmetric compactifications provide a
promising setting for obtaining realistic models of particle
physics, a microscopic interpretation of black hole entropy in
string theory is best understood for supersymmetric black holes,
and various kinds of supersymmetric solutions have transformed our
understanding of quantum field theory via the AdS/CFT
correspondence and its generalisations.

Here we would like to review the remarkable progress that has been
made over the past two and a half years in classifying such
solutions. Recall that if one sets all of the matter-fields
(fluxes) to zero, bosonic supersymmetric solutions must have
metrics with special holonomy. Thus, the basic problem is to find
the appropriate generalisation for more general solutions with
non-vanishing fluxes. The key mathematical tools
\cite{Gauntlett:2002sc,11dclass} are ``$G$-structures". The rough
idea is to maintain the tensors arising in special holonomy
manifolds, but to relax the differential conditions that they
usually satisfy. For example, a Calabi-Yau three-fold has a
K\"ahler form and a $(3,0)$ form which satisfy certain algebraic
conditions. The necessary and sufficient conditions for a manifold
admitting such tensors to have $SU(3)$ holonomy are that both of
these forms are closed. More general $SU(3)$-structures arise by
relaxing these differential conditions in a precise way.

$G$-structures have now been used to classify various kinds of
supersymmetric solutions of different supergravity theories. One
of the purposes of this article is to summarise what has been
achieved and to highlight some remaining issues that could be
examined further. Broadly speaking the work completed falls into
three categories. Firstly, a classification of the most general
supersymmetric solutions of supergravity theories arising as the
low-energy limit of string/M-theory. Secondly, a similar
classification for simpler supergravity theories in
lower-dimensions. And thirdly, a classification of specific
classes of solutions of physical interest. For example, solutions
that are products, possibly warped products, of four-dimensional
Minkowski space with a six- or seven-dimensional compact internal
space are of interest for phenomenological reasons, while similar
products of anti-de-Sitter ($AdS$) space with a compact internal
space are of interest for the $AdS$/CFT correspondence.

Having obtained such a classification one can use the results to
try and construct new solutions in explicit form. This has also
been a very successful endeavour. Four prominent examples are: (i)
the discovery of the maximally supersymmetric G\"odel solution of
D=5 supergravity, which initiated investigations into the role of
closed timelike curves in string/M-theory \cite{5dclass} (ii) the
construction of a very rich class of geometries corresponding to
1/2 BPS excitations in $AdS$/CFT dualities \cite{llm} (iii) the
discovery of an infinite class of new Sasaki-Einstein manifolds
$Y^{p,q}$ that provide new $AdS_5\times Y^{p,q}$ solutions of type
IIB supergravity \cite{m6,se} and (iv) the discovery of
supersymmetric black rings in $D=5$ supergravity, i.e., black
holes with horizons of topology $S^1\times S^2$ rather than the
usual $S^3$
\cite{Elvang:2004rt,Gauntlett:2004wh,Bena:2004de,Elvang:2004ds,Gauntlett:2004qy}.

The discovery of the infinite class of Sasaki-Einstein manifolds
$Y^{p,q}$ was recently reviewed in \cite{Gauntlett:2004hs} and so
we shall not discuss them much here. However, since that review
appeared there has been some very interesting developments:
following some further studies of the geometry
\cite{Martelli:2004wu} the corresponding dual conformal field
theories have now been identified \cite{Benvenuti:2004dy} and,
using the techniques of ``a-maximisation"
\cite{Intriligator:2003jj}, the central charges of the conformal
field theories have been computed
\cite{Martelli:2004wu,Bertolini:2004xf,Benvenuti:2004dy} and shown
to agree precisely with those predicted in \cite{se} from a
calculation of the volumes of the $Y^{p,q}$. It is worth
emphasising that before the discovery of the $Y^{p,q}$ in
\cite{m6,se} and the work of \cite{Benvenuti:2004dy} there were
only a few $AdS_5$/CFT examples\footnote{Ignoring orbifolds of
these examples, which are certainly interesting.} where both the
geometry was explicit and the field theory was identified.
Additional recent work in this area appears in
\cite{Herzog:2004tr,Benvenuti:2004wx}.

In this article we will briefly review the construction of the
supersymmetric black ring solutions. The surprising discovery of
black ring solutions was first made in a non-supersymmetric
setting in $D=5$ by Emparan and Reall in \cite{Emparan:2001wn}.
The $S^1\times S^2$ topology of the event horizon is prevented
from collapsing by the rotation of the black ring. The
supersymmetric generalisation, for minimal D=5 supergravity, was
found in \cite{Elvang:2004rt} using the classification of
solutions for this theory carried out in \cite{5dclass}. This was
further generalised to black ring solutions of D=5 supergravity
coupled to vector multiplets in
\cite{Bena:2004de,Elvang:2004ds,Gauntlett:2004qy} and
multi-concentric rings were found in
\cite{Gauntlett:2004wh,Gauntlett:2004qy}. An interesting class of
solutions occurs when there are two vector multiplets as this
theory arises from a toroidal reduction of string/M-theory. In
this case the black ring solutions are specified by seven
parameters and yet only carry five independent conserved charges.
In other words, in marked contrast to four-dimensional black
holes, these black rings violate a naive generalisation of the
powerful uniqueness theorems for four-dimensional black holes.
This non-uniqueness is a striking feature of the supersymmetric
black rings and it will be very interesting to see how it is
incorporated in a microscopic identification of the black hole
entropy. Some analysis of such an identification appears in
\cite{Bena:2004tk,Cyrier:2004hj}.

\section{Classifying Supergravity Solutions}

Consider a general supergravity theory whose bosonic fields
consist of a metric, $g_{\mu\nu}$, and some possible matter
fields, often referred to as ``fluxes", which are generically a
set of $p$-form potentials for various $p$. We are only interested
in bosonic solutions and so we set all of the fermions to zero.
The equations of motion for such configurations are then,
schematically, \bea
R_{\mu\nu}-\frac{1}{2}Rg_{\mu\nu}=T_{\mu\nu}\nn Matter Equations,
\eea where $T_{\mu\nu}$ is the energy momentum tensor of all the
matter fields and the second line refers to the equations of
motion for the matter fields and also Bianchi identities. A
supersymmetric solution is one that is left invariant under
certain supersymmetry transformations. Since the fermions are zero
the supersymmetry variation of the bosonic fields is automatically
zero and so we just need to ensure that the variation of the
fermionic fields are zero. This requires that the solutions admit
``Killing spinors" $\epsilon$ that satisfy \be\label{condssusy}
\hat\nabla_\mu\epsilon=0,\qquad M\epsilon=0. \ee Here $\hat\nabla$
is a connection that is schematically of the form
$\hat\nabla=\nabla +fluxes \cdot \gamma$ where $\nabla$ is the
ordinary Levi-Civita connection and the remaining pieces are
matter fields contracted with various anti-symmetrised products of
gamma-matrices denoted generically by $\gamma$. In mathematical
terms $\nabla$ is a connection on the spin bundle while, in
general,  $\hat\nabla$ is a connection on the Clifford bundle. In
\p{condssusy} $M$ is a matrix, that is sometimes present, that
depends on the fluxes and gamma matrices and leads to additional
algebraic conditions.

Now let us recall the situation when all of the matter fields
(fluxes) are zero. The problem then boils down to solving\be
R_{\mu\nu}=0,\qquad \nabla_\mu\epsilon=0,\ee i.e. Ricci-flat
manifolds with covariantly constant spinors. Actually on a
Euclidean manifold, as in a compactification manifold, the latter
implies the former, but not, in general, on a Lorentzian manifold.
Manifolds with covariantly constant spinors have special holonomy.
In the familiar Euclidean case, Berger's classification implies
that the manifold can be Calabi-Yau, with $SU(n)$ holonomy in $2n$
dimensions, hyper-K\"ahler, with $Sp(n)$ holonomy in $4n$
dimensions, have $G_2$ holonomy in seven-dimensions and Spin(7)
holonomy in eight dimensions. The Lorentzian case is less studied.
However, some possibilities relevant for string/M-theory are
discussed in \cite{bryant,Figueroa-O'Farrill:1999tx}. For example,
in D=11 supergravity, the general solutions have $SU(5)$ or
$Spin(7)\ltimes \bR^9$ holonomy.

The question therefore arises as to how to generalise this
classification when the fluxes are non-zero. The key
generalisation of special holonomy that we employ
\cite{Gauntlett:2002sc,11dclass} is a $G$-structure, as we discuss
in the next section. Before doing that we mention that an
alternative approach, analysing the ``generalised holonomy" of the
connection $\hat\nabla$ appearing in \p{condssusy}, has been
advocated in \cite{genhol}. However, as discussed in
\cite{Gauntlett:2004hs}, this approach, so far, misses some
information contained in the conditions for supersymmetry, but it
is possible that it could be developed further. We also mention
that an effective approach for constructing ansatze for specific
sub-classes of solutions was discussed in \cite{warner}: in fact
this method can also be viewed from a $G$-structure point of view.

\section{$G$-Structure Classification}

\subsection{$G$-structures}
Let us begin with the abstract definition (see e.g. \cite{joyce}).
Consider an $n$-dimensional manifold $M$. The set of all frames
defines the frame bundle $F(M)$ which is a principal $Gl(n)$
bundle, consistent with the fact that an element of $Gl(n)$ will
transform one frame into another frame. A $G$-structure is then
defined as a principal $G$ sub-bundle of $F(M)$.

In cases of interest, this definition is entirely equivalent to
the existence of no-where vanishing tensors. For example, a
Euclidean metric $g_{ab}$ on an $n$-dimensional manifold is
equivalent to an $O(n)$ structure as one can use the metric to
restrict to orthonormal frames. Similarly, if we supplement this
with an orientation $\epsilon_{a_1\dots a_n}$ we can restrict to
orthonormal frames with a particular orientation and this is the
data required for an $SO(n)$ structure. If we are in
even-dimensions, $n=2m$, with an almost complex structure
${J_a}{}^b$, satisfying ${J_a}{}^b {J_b}{}^c=-\delta^c_a$, this
defines a $Gl(m,\bC)$ structure. If we also have a hermitian
metric this defines a $U(m)$ structure. If in addition there is an
$(m,0)$-form $\Omega$ we have an $SU(m)$ structure. While these
structures are somewhat familiar, more exotic $G$-structures such
as $Spin(7)\ltimes \bR^9 \subset SO(1,10)$ also appear in
classifying supergravity solutions as we shall mention later. An
important feature of a $G$-structures is that it allows any tensor
to be decomposed into representations of $G$.

It is important to emphasise that the existence of a $G$-structure
is topological and does not entail any differential conditions on
the tensors. For example, the existence of a $Gl(m,\bC)$
structure, which we noted above is equivalent to the existence of
a no-where vanishing almost complex structure, does not imply that
the manifold is complex, which requires additional differential
conditions to be satisfied  (that the Nijenhuis tensor vanishes).

That being said, there is a natural way to classify $G$-structures
using differential conditions satisfied by the tensors
encapsulated in the ``intrinsic torsion" of the $G$-structure. For
illustration consider a $G$-structure with $G\subset SO(n)$. In
this case we always have a metric and hence a Levi-Civita
connection $\nabla$. Roughly, one takes the covariant derivative
of all the tensors $\eta$ defining the $G$-structure and then
decomposes the result into irreducible $G$-modules $W_i$ acting on
$\eta$ to obtain the intrinsic torsion: \be \nabla\eta\to T\in
\oplus W_i\cong\Lambda^1\otimes g^\perp. \ee Here $\Lambda^1$ is
the space of one-forms and $g\oplus g^\perp=so(N)$ where $g$ is
the Lie algebra of $G$. More precisely, and to explain the last
expression, we use the fact that there is no obstruction to
finding a connection $\nabla'$ such that $\nabla'\eta=0$ and hence
$\nabla\eta=(\nabla-\nabla')\eta$. Define the con-torsion tensor
$(\nabla-\nabla')$ and recall that the con-torsion has the same
information as the torsion. We next note that $(\nabla-\nabla')\in
\Lambda^1\otimes so(n)$. However, since $\eta$ is $G$-invariant,
only the part of the con-torsion in $\Lambda^1\otimes g^\perp$
acts on $\eta$ and this is clearly the part of the con-torsion
that is independent of the choice of $\nabla'$. This part is
called the intrinsic con-torsion. The intrinsic con-torsion, or
equivalently the intrinsic torsion $T$, can then be decomposed
into the $G$-modules $W_i$. For a more detailed discussion in the
physics literature see \cite{Gauntlett:2004hs} and for many
explicit examples see \cite{int-tor}.

The most extreme case when all $W_i=0$, i.e. vanishing intrinsic
torsion, is equivalent to $\nabla\eta=0$ which is equivalent to
having special holonomy $G$. Thus, the intrinsic torsion is a
precise measure of the deviation away from special holonomy $G$
and it is for this reason that $G$-structures are useful in
classifying supergravity solutions when the fluxes are
non-vanishing.

\subsection{Classifying supergravity solutions}

The classification of supersymmetric supergravity solutions has
three main steps. The first two steps analyse the information
contained in \p{condssusy}; the first is purely algebraic, while
the second is also differential. The third step is to impose the
additional conditions that the equations of motion are satisfied.

1. The first observation is that the Killing spinor $\epsilon$
defines a canonical $G$-structure\footnote{See e.g.
\cite{Gauntlett:2004hs} for some discussion on the issue of
whether or not this $G$-structure is globally defined.}. Indeed
the isotropy group of the spinor $G\subset Spin(d)$ (or $\subset
Spin(1,d-1)$) is in fact a $G$-structure. In the language above,
the tensors defining the $G$-structure can be simply constructed
from $\epsilon$ as bi-linear differential forms, roughly of the
form $\bar\epsilon\gamma_{(n)}\epsilon$ where $\gamma_{(n)}$ is a
basis of the Clifford algebra consisting of anti-symmetrised
products of $n$ gamma-matrices. These tensors will satisfy a
number of algebraic conditions corresponding to the $G$-structure.
They can be obtained, for example, by doing various Fierz
transformations.

2. The second step utilises the information that the Killing
spinor $\epsilon$ satisfies the differential, and possibly
algebraic, conditions in \p{condssusy}. After a detailed analysis
one finds that the intrinsic torsion of the $G$-structure is
restricted and that the fluxes are correlated with the intrinsic
torsion. In general there can be components of the flux which drop
out of the supersymmetry conditions \p{condssusy} completely.

3. By analysing the integrability conditions for the Killing
spinor equations \p{condssusy}, one can show that some but not all
of the equations of motion are automatically satisfied
\cite{wsh,11dclass}. In general, it is sufficient to impose just
the Bianchi identities and the equations of motion for the matter
fields, and, in the Lorentzian case, at most one component of the
Einstein's equations \cite{11dclass}.

Typically, this classification\footnote{A complete classification
of all supersymmetric solutions for all supergravity theories in
{\it explicit} form is, of course, well beyond the scope of
current techniques. For example, just consider special holonomy
manifolds.} provides the most general local form of the solution
in terms of $G$-structure data (sometimes of a lower-dimensional
manifold) as well as a number of differential conditions that
remain to be solved. These techniques have now been applied in a
number of different contexts. The work falls into three broad
classes which we now discuss in the subsequent sections.

\section{The Most General Solutions for String/M-Theory}

The techniques described above have been used to classify the most
general supersymmetric solutions of D=11 supergravity in
\cite{11dclass,11dnull}. Recall that the bosonic fields of D=11
supergravity consist of the metric plus a four-form field
strength. The most general supersymmetric solution will preserve
(at least) one supersymmetry, i.e. admit a Killing spinor
$\epsilon$. It turns out that there are two distinct classes of
supersymmetric solutions, one with an $SU(5)$ structure and the
other with a $Spin(7)\ltimes \bR^9$ structure, corresponding to
the two possible isotropy groups of spinors of $Spin(10,1)$
\cite{bryant}. The two cases can be distinguished by the algebraic
conditions satisfied by the bi-linears constructed from the
spinor. From a single spinor we can construct a one-form, $K$, a
two-form $\Omega$, and a five-form $\Sigma$. In particular, if $K$
is time-like we have an $SU(5)$ structure, while if $K$ is null we
have a $Spin(7)\ltimes \bR^9$ structure. It turns out that in both
cases the Killing spinor equation implies that $K$ is Killing.

The time-like case has an $SU(5)$ structure in eleven-dimensions.
This structure can be demystified by noting that the time-like
one-form $K$ specifies a ten-dimensional base manifold $M$ and
that (essentially) the two-form $\Omega$ and $\chi$, the part of
the five-form $\Sigma$ independent of $K$, satisfy the algebraic
conditions of a more familiar ten-dimensional $SU(5)$ structure on
$M$. The most general local form of the metric is given by \be
ds^2=-\Delta^2(dt+\omega)^2+\Delta^{-1}ds^2(M_{10}), \ee where
$\Delta$ is a function and $\omega$ is a one-form on the base
manifold $M_{10}$. The intrinsic torsion of the $SU(5)$-structure
on $M_{10}$ is only weakly constrained. Such structures have five
torsion classes $W_i$ (see e.g. \cite{int-tor}). The only
constraint here is that the one-form $\chi\lrcorner d\chi$, which
specifies $W_5$, must be equal to $12 \Delta^{-1}d\Delta$. The
components of the four-form are almost determined by this data,
but there is a component that isn't. To ensure that we have a
solution to the equations of motion it is sufficient to impose the
Bianchi identities and equations of motion for the four-form,
which leads to elaborate differential conditions on
$\Delta,\omega$ and the structure that still need to be solved.
Note, in particular, that they constrain the part of the flux not
fixed by the Killing spinor equation alone.

 When $K$ is null there is a
$Spin(7)\ltimes \bR^9$ structure. Roughly, this corresponds to
being able to choose a null frame $e^+,e^-,e^i,e^9$, with
$i=1,\dots,8$, with a $Spin(7)$ structure, specified by a Cayley
four-form, constructed from the $e^i$ only. For more details on
this structure and the analysis of the Killing spinor equation
leading to the most general local form of the solution, we refer
to \cite{11dnull}.

It is satisfying that such a complete description of the geometry
underlying the most general supersymmetric solutions can be
obtained. The final result is rather general: for example in the
timelike case we noted that the geometry involves an $SU(5)$
structure in $D=10$ with only weakly constrained intrinsic
torsion. But in fact this is what one might expect since the
result contains all possible supersymmetric solutions preserving
(at least) one time-like Killing spinor. Nevertheless, it is still
useful in constructing new explicit solutions as shown in
\cite{11dclass}.

In most applications we are interested in $D=11$ supersymmetric
solutions preserving more than one supersymmetry, and hence it is
of interest to refine the classification and determine the
conditions imposed by the presence of more supersymmetries. The
classification of solutions preserving all supersymmetries in
$D=10$ and $D=11$ has already been carried out some time ago
\cite{Figueroa-O'Farrill:2002ft}, but the techniques used apply
only to this case. However, a refinement using $G$-structures is
possible. The preservation of more than one supersymmetry reduces
the structure groups $SU(5)$ or $Spin(7)\ltimes \bR^9$ further and
this is a useful tool in refining the classification. Indeed
following the suggestion in \cite{11dclass} and the work
\cite{oisin} this approach is being pursued in
\cite{Gillard:2004xq,Cariglia:2004ym,Cariglia:2004wu}. It is clear
that much progress can be made in this direction. It should be
noted, however, that there will be many cases which will have an
identity structure and the technology of $G$-structures will then
not be particularly helpful in organising the calculation.

It would be very interesting if a detailed understanding of
supersymmetric solutions preserving, for example, more than 1/2
supersymmetry can be obtained. New solutions, supplementing the
known examples of G\"odel spacetimes and pp-waves, are likely to
have interesting applications in M-theory. An interesting result
is that backgrounds with more than 24 supersymmetries are locally
homogeneous \cite{Figueroa-O'Farrill:2004mx}.

A similar analysis for type IIB supergravity is desirable and some
nice results using $G$-structure techniques have just appeared in
\cite{jan}.

\section{Lower-Dimensional Supergravities}

Similar techniques have also been used to study supergravity
theories in lower dimensions. The major motivation for studying
these theories is that they provide powerful ways of obtaining
solutions relevant to string/M-theory after uplifting them to
D=10/D=11 supergravity. The simplicity of these supergravity
theories, compared to those in D=10 and 11, allows one to be much
more explicit about the geometry. The first analysis of this kind
was first carried out by Tod in the context of four-dimensional
supergravity \cite{Tod1,Tod2}. Here we will illustrate the
analysis with a discussion of minimal D=5 supergravity
\cite{5dclass}. The results have led to interesting new classes of
explicit solutions including the surprising discovery of the
maximally supersymmetric G\"odel spacetime and the supersymmetric
black rings, the latter to be reviewed in section 7.

\subsection{D=5 Minimal Supergravity}

The bosonic fields of $D=5$ minimal supergravity consist of a
metric and a vector potential with field strength $F$. A bosonic
solution to the equations of motion is supersymmetric if it admits
a super-covariantly constant spinor obeying \be\label{kspin}
 \left[ D_\alpha + \frac{1}{4\sqrt{3}}  \left(
 \gamma_\alpha{}^{\beta\gamma} - 4 \delta^{\beta}_{\alpha}\gamma^{\gamma}
  \right){}F_{\beta \gamma} \right] \epsilon^a = 0.
\ee where $\epsilon^a$ is a commuting symplectic Majorana spinor.

{}From a single commuting spinor $\epsilon^a$ we can construct a
scalar $f$, a 1-form $V$ and three 2-forms $X^{(a)}$. The Clifford
algebra implies that these tensors satisfy various algebraic
conditions corresponding to a $G$-structure. As in $D=11$
supergravity there are two types of supersymmetric solutions to
consider. When $V$ is time-like, $f,V,X$ specify an $SU(2)$
structure, while when $V$ is null they specify an $\bR^3$
structure.

The next step in the classification programme is to analyse the
differential conditions imposed on these structures arising from
\p{kspin}. After a detailed analysis one derives a local form of
the most general supersymmetric solutions. Let us summarise the
conditions for the time-like case. We find that the metric can be
written as \be ds^2 = -f^2(dt+\omega)^2 + f^{-1} ds^2(M_4) \,,
\label{metric} \ee where $M_4$ is an arbitrary hyper-K\"ahler
space (this is the analogue of $M_{10}$ having a certain $SU(5)$
structure in the $D=11$ case), and $f$ and $\omega$ are a scalar
and a one-form on $M_4$, respectively. The two-form field strength
is given by \be F=\frac{\sqrt 3}{2}d[f(dt+\omega)]-\frac{1}{\sqrt
3}G^- \ , \ee where $G^{\pm}\equiv \frac{1}{2}f(d\omega\pm
*d\omega)$, with $*$ the Hodge dual on $M_4$. In order that all of
the equations of motion are satisfied $f$ and $\omega$ must
satisfy \be d G^+ =0 \, , \qquad \Delta f^{-1} = \frac{4}{9}
(G^+)^2 \,, \label{eom} \ee where $\Delta$ is the Laplacian on
$M_4$. We refer to \cite{5dclass} for the analogous result for the
null case.

This formalism has been effectively used to construct many new
solutions. One important point to emphasise is that we can obtain
sensible solutions from pathological hyper-K\"ahler base spaces.
For example, we can obtain $AdS_2\times S^3$ using a singular
Eguchi-Hanson type base, while the G\"odel solution can be
constructed from a singular negative mass Taub-NUT base space.

The black ring solutions, to be reviewed later, have base space
$\bR^4$. For such a base the above equations are actually linear
if they are solved in the correct order and this can be used to
construct the black ring solutions \cite{Bena:2004de}. We adopt a
different and very effective approach \cite{5dclass} by first
writing the metric for $\bR^4$ in Gibbons-Hawking form: \bea
\label{ghflat}
ds^2&=&H[dr^2+r^2(d\theta^2+\sin^2(\theta)d\phi^2)]
+H^{-1}(d\psi+\cos\theta d\phi)^2 \eea with $H=1/r$. We will
further demand that the tri-holomorphic vector field
$\partial_\psi$ is a Killing vector of the five-dimensional
metric. The significance of this is that the most general solution
is then specified by three further harmonic functions, $K$, $L$
and $M$ on $\bR^3$ (with coordinates $(r,\theta,\phi)$). In
particular the general solution has \bea\label{ghflateqs}
f^{-1}&=&H^{-1}K^2+L \nn
\omega&=&(H^{-2}K^3+\frac{3}{2}H^{-1}KL+M)(d\psi+\cos\theta d\phi)
+\hat\omega_i dx^i \eea with $\bomega$ obtained by solving
$\nabla\times \hat\bomega=H\nabla M-M\nabla H+\frac{3}{2}(K\nabla
L-L\nabla K)$.

\subsection{Generalisations}

Similar classifications have now been carried out for a number of
different supergravity theories in four
\cite{Tod1,Tod2,4dcald1,4dcald2}, five
\cite{5dclass,5dgauged,Gutowski:2004yv}, six \cite{6dQMW,6dcam}
and seven \cite{7d} dimensions. Both gauged and ungauged theories
have been considered as well as the possibility of including
various matter fields. Although these investigations are
extensive, they are not exhaustive and further theories could be
considered. Furthermore, while we understand the cases of
preservation of minimal and maximal supersymmetry (see also
\cite{fig,fiol,ortin1}), we would like to refine the
classification to consider the cases in between, just as in the
$D=11$ case. An investigation for the case of $D=7$ has been
initiated in \cite{oisin}.

It should  also be profitable to continue to seek new explicit
solutions using the formalism. For example, an interesting class
of asymptotically $AdS$ black holes were found in gauged
supergravity in \cite{Gutowski:2004ez,Gutowski:2004yv}. Following
the discovery of black rings, it seems plausible that
asymptotically $AdS$ black rings also exist. It would also be very
interesting to construct supersymmetric black holes with other
non-spherical topologies, or alternatively prove that they do not
exist by generalising the results in
\cite{Reall:2002bh,Gutowski:2004bj}.

\section{Compactifications for Phenomenology and AdS/CFT}

$G$-structures can also be used to classify particular classes of
geometries of physical interest in string theory. For example, one
can study warped compactifications from D=10,11 to $d=4$ Minkowski
spacetime ($\bM_4$) for model building, or to $AdS$ spaces for
AdS/CFT applications. This area was recently reviewed in
\cite{Gauntlett:2004hs} which contains further discussion.

A warped compactification has a metric of the form \be ds^2=
e^{2\Delta}ds^2(M_d)+ds^2(X)\ee where $ds^2(M_d)$ is a metric on
$\bM_d$ or $AdS_d$, $\Delta$ is function on the internal space $X$
with metric $ds^2(X)$. The significance of this ansatz is that it
preserves all of the isometries of $M_d$. An interesting
observation is that if we write the $AdS_d$ metric in Poincar\'e
coordinates: \be ds^2=e^{-2mr}ds^2(\bM_{d-1})+dr^2\ee then,
locally, we can view the $AdS_d$ cases as special examples of the
warped compactifications to $\bM_{d-1}$ by thinking of the radial
coordinate as a coordinate on the internal space.

Studies of compactifications to $\bM_4$ with fluxes were initiated
long ago in \cite{strom,hull}. More recently these results were
rederived using $G$-structures in \cite{Cardoso:2002hd,int-tor}.
General compactifications to $\bM_4$ with fluxes have been studied
using $G$-structures for D=11 in
\cite{Behrndt:2003uq,kmt,Behrndt:2003zg,Lukas:2004ip,Behrndt:2004bh},
for type IIB in \cite{Dall'Agata:2004dk,Frey:2004rn} and for
massive IIA in \cite{Behrndt:2004km}. Investigations into mirror
symmetry were carried out in \cite{Gurrieri:2002wz}. In this area,
it is too much to hope for explicit compact
solutions\footnote{Non-compact solutions have important
applications in gauge/gravity dualities as in e.g.
\cite{Maldacena:2000yy}. See \cite{int-tor}.}. After all recall
that no explicit compact Calabi-Yau three-folds are known.
Progress in the study of Calabi-Yau manifolds stem from the
powerful results on the existence of Calabi-Yau metrics arising
from Calabi's theorem. It would be most desirable to have
analogous existence theorems for the geometries arising when the
fluxes are non-vanishing building on the $G$-structure results. It
seems likely that this will involve hard analysis.

On the other hand the classification of $AdS$ compactifications
has been very fruitful for constructing new explicit solutions.
The classification of the most general $AdS_5$ geometries in D=11
supergravity produced an infinite number of new solutions
\cite{m6}. Furthermore, after dimensional reduction and T-duality,
a sub-class led to the discovery of an infinite new set of
Sasaki-Einstein manifolds $Y^{p,q}$ which give rise to IIB $AdS_5$
solutions \cite{se}. Moreover these were generalised to arbitrary
odd dimensions in \cite{gse}, and generalised further in
\cite{Gauntlett:2004hs,Chen:2004nq}, the $D=7$ examples giving new
$AdS_4$ solutions in $D=11$.

It seems likely that similar general classifications of $AdS_5$
geometries in type IIB or $AdS_4$ in D=11 or type IIB will lead to
the discovery of new classes of explicit solutions. A study for
$AdS_4$ in type II appears in \cite{Lust:2004ig} and $AdS_3$ in
$D=11$ appears in \cite{Martelli:2003ki}. Refining such
classifications should also be a profitable undertaking. The
refinement of the classification of the $AdS_5$ geometries arising
in $D=11$ supergravity to have $N=2$ supersymmetry was presented
in \cite{llm}. Interestingly, the result in \cite{llm} was
obtained by analytically continuing an ansatz for a class of
solutions describing 1/2 BPS excitations in $AdS_7\times S^4$ and
$AdS_4\times S^7$. A similar analysis of 1/2 BPS excitations of
$AdS_5\times S^5$ was also undertaken and a very rich set of
explicit solutions were obtained. Generalising these results to
situations with less supersymmetry may also be rewarding.

\section{D=5 Supergravity and Black Rings}

We now review the construction of the new supersymmetric black
ring solutions.

\subsection{Minimal D=5 supergravity}
The single supersymmetric black ring solution of minimal $D=5$
supergravity was found \cite{Elvang:2004rt} by directly solving
the equations \p{eom} for a flat base space $\bR^4$. The solution
was rederived in \cite{Gauntlett:2004wh} by writing $\bR^4$ in
Gibbons-Hawking form, \p{ghflat}, and then using equations
\p{ghflateqs}. This immediately leads to a straightforward
construction of multi-concentric black rings as we now discuss.

The three harmonic functions $K,L$ and $M$ in \p{ghflateqs} for a
single back ring can all be expressed in terms of a single
harmonic function $h_1$ on $\bR^3$ given by \be h_1=\frac{1}{|{\bf
x}-{\bf x_1}|} \ee with a single centre on the negative $z$-axis:
${\bf x}_1\equiv (0,0,-R^2/4)$. Specifically: \bea
K=-\frac{q}{2}h_1,\qquad L=1+\frac{Q-q^2}{4}h_1,\quad
M=\frac{3q}{4}-\frac{3qR^2}{16}h_1 \eea and it is simple to obtain
an expression for $\omega$ in \p{ghflateqs}. The solution is
asymptotically flat\footnote{The generalisation that is
asymptotically G\"odel appears in \cite{ortin2}.}. It appears
singular at ${\bf x}={\bf x}_1$ and ${\bf x}={\bf 0}$ (recall
$H=1/|{\bf x}|$ in \p{ghflat}), but both of these are coordinate
singularities. The key to the geometry lies at ${\bf x}={\bf x}_1$
which turns out to correspond to the event horizon of the black
ring with topology $S^1\times S^2$. The radius of the $S^2$ is
$q/2$ and that of the $S^1$ is $l$ defined by \be \label{defnofl}
l\equiv \sqrt{ \frac{3(Q-q^2)^2}{4q^2} - 3R^2  } \ . \ee Reality
of $l$ ensures that the geometry is free from closed-time-like
curves \cite{Elvang:2004rt}.

The solution depends on three parameters $q$, $Q$ and $R$ which
are uniquely specified by the three independent conserved charges:
the mass and two angular momentum $J_1$ and $J_2$ (in five
dimensions black holes can rotate in two two-planes) with
$J_1-J_2\propto qR^2$. Supersymmetry implies that the total
electric charge is not an independent charge but is specified by
the mass. Note that this uniqueness property is not shared by the
black ring solutions of the more general supergravity theories
discussed in the next sub-section.

It is interesting to observe \cite{Elvang:2004rt} that upon
setting $R=0$, so that all four harmonic functions $H,K,L$ and $M$
have the same single centre at ${\bf x}={\bf 0}$,  we recover the
black hole solution of \cite{bmpv,Gauntlett:1998fz} which has a
horizon with $S^3$ topology and $J_1=J_2$.

Having found the single black ring solution in this language, the
construction of multi concentric black rings is very simple
\cite{Gauntlett:2004wh}. The three harmonic functions $K,L$ and
$M$ are now multi-centred: \bea\label{mbra} K=-\frac{1}{2}
\sum_{i=1}^N q_i h_i,\quad L=1+ \frac{1}{4} \sum_{i=1}^N (Q_i -
q_i{}^2) h_i,\quad M= \frac{3}{4} \sum_{i=1}^N q_i - \frac{3}{4}
\sum_{i=1}^N q_i |{\bf{x}}_i|h_i\eea with $h_i=1/|{\bf x}-{\bf
x}_i|$. The solution is parametrised by ${\bf x}_i$, $Q_i$ and
$q_i$. The solution is fully specified after solving for $\omega$
in \p{ghflateqs}, which is not straightforward in general.
However, it is straightforward when all of the poles are located
on the $z$-axis, when the solution inherits another Killing
symmetry. In general, the apparent singularities at the centres
${\bf x}_i$ each correspond to Killing horizons with topology
$S^1\times S^2$. Thus the solutions correspond to multi black
rings. For the $ith$ ring, the radius of the $S^2$ is $q_i/2$ and
that of the $S^1$ is $l_i(Q_i,q_i,{\bf x}_i)$, the obvious
generalisation of \p{defnofl}.

The solutions are invariant under the action of $\partial_\psi$
and the $S^1$ direction of each horizon lies on an orbit of this
vector field. To see that the solutions describe concentric black
rings, it is helpful to consider the orbits of $\partial_\psi$ in
$\bR^4$, which, roughly, can be thought of as the $\bR^4$ at
asymptotic infinity in the solution. In particular, the location
of the pole in $\bR^3$ for a given ring corresponds to a two-plane
in $\bR^4$ in which the $S^1$ of the ring lies. Moreover, all of
these $S^1$s are concentric. In the simple case that all of the
poles lie in a given direction, for example the negative $z$-axis,
all of the $S^1$s lie in the same two-plane. While if the poles
lie on a single line in $\bR^3$, for example the $z$-axis, then
the black rings lie in one of two orthogonal two-planes. Finally,
our solutions also allow for the possibility of a single black
hole being located at the common centre of all of the rings, by
simply choosing one of the centres ${\bf x}_i$ to be ${\bf 0}$. It
is an interesting open question as to whether or not there are
more general multi-black ring solutions that are not concentric.

By restricting to solutions with all of the poles located along
the $z$-axis, we can analyse the geometry in more detail.
Consider, therefore, the general solution \p{mbra} with ${\bf
x_i}=(0,0,-k_iR_i^2/4)$ and $k_i=\pm 1$. One finds that in order
to eliminate Dirac-Misner strings, which would lead to closed
time-like curves, it is necessary to impose
$\Lambda\equiv\Lambda_1=\Lambda_2=...$, where $\Lambda_i\equiv
(Q_i-q_i^2)/2q_i$. In this case, the radii of the rings are given
by \be l_i \equiv \sqrt{3 \left[\Lambda^2 - R_i^2 \right] } \ .
\ee This expression implies that we can place the poles anywhere
on the $z$-axis provided that $R_i^2<\Lambda^2$. It is interesting
to observe that as the location of the pole goes to larger values
of $|z|$, i.e. as $R_i$ increases, the circumference of the rings
get uniformly smaller, perhaps contrary to one's intuition.

Another interesting observation is that there are configurations
of two (say) black rings with the same asymptotic charges as the
single black hole solution of \cite{bmpv,Gauntlett:1998fz}. For
example, by taking one pole to lie on the positive $z$-axis and
another to lie on the negative $z$-axis one can obtain $J_1=J_2$,
just as for the black hole. Moreover, and surprisingly, it is
possible to choose the parameters such that the sum of the areas
of the horizons of the two black rings is greater than, less than,
or equal to the area of the black hole. In particular, it is
possible for the black rings to be entropically preferred.

This should be contrasted with supersymmetric black holes with
spherical topology. For example, in $D=4$, extreme
Reissner-Nordstrom black holes with $M=|Q|$ have areas which scale
as $M^2$. Since $M_1^2+M_2^2< (M_1+M_2)^2$ the entropy of the
multi-black holes is less than that of a single black hole with
the same total charge and mass. It will be interesting to see how
the contrasting property of the rings is accounted for from a
microscopic state counting point of view.

\subsection{D=5 supergravity coupled to vector multiplets}

It is interesting to generalise the black ring solutions just
described to solutions of minimal $D=5$ supergravity coupled to an
arbitrary number of vector multiplets as this gives important new
solutions in string/M-theory. Recall, for example, that these
supergravity theories arise as part of the low-energy effective
action of M-theory reduced on a Calabi-Yau three fold. An
interesting special case is the so-called $STU$ model which has
two vector multiplets and hence three vector fields in total. This
model can arise from M-theory reduced on a six-torus and also from
type IIB supergravity reduced on a five-torus. In the latter
context, the three-charge black hole of this model (with spherical
topology) then uplifts to a $D1,D5$, momentum system which is the
setting for a precise understanding of the black hole entropy
\cite{Strominger:1996sh,bmpv}.

The construction of the single black ring solution of these more
general supergravity theories was carried out in
\cite{Bena:2004de,Elvang:2004ds,Gauntlett:2004qy}. The
construction pursued in \cite{Gauntlett:2004qy} exactly mimics the
construction described above for the minimal theory. One first
utilises the classification of the most general supergravity
solutions carried out in \cite{Gutowski:2004yv}. Actually, gauged
supergravity was considered in \cite{Gutowski:2004yv} but the
results can easily be adapted to the ungauged case of interest
here. As for minimal supergravity, there are timelike solutions
and null solutions, and the black rings lie in the time-like case.
The time-like solutions again have a hyper-K\"ahler base space and
in the special case that it is of Gibbons-Hawking form, it was
shown in \cite{Gauntlett:2004qy} that, for the case of minimal
supergravity coupled to $n-1$ vector multiplets, the solution is
determined by $2n+2$ harmonic functions in $\bR^3$ (one of which
specifies the hyper-K\"ahler manifold).

Using these results one constructs the black ring solution exactly
as before using a flat hyper-K\"ahler base space. Furthermore, it
is simple to generalise this to multi-concentric black rings with
the optional possibility of a black hole (of the type
\cite{bmpv,chsab}) at the centre.

Let us close with a brief comment about the single black ring
solution of the $STU$ model. It is specified by
 seven parameters $Q_i$, $q_i$ and $R$, where $i=1,2,3$ is an index for the three
 vector fields present in the model. On the other hand the solution only carries five
 independent conserved charges: three electric charges (which determine the mass)
 and two angular momentum. Thus, in contrast to
the single ring of the minimal theory, the solution is not
uniquely determined by the conserved charges. It will be
interesting to see how this property is accounted for from a
microscopic point of view. An important observation is that the
extra parameters correspond to dipole charges carried by the
system and this leads to an identification of the system in terms
of branes and dipole branes \cite{Bena:2004de,Elvang:2004ds} and
to a microscopic identification of the entropy 
\cite{Cyrier:2004hj} (see also \cite{Bena:2004tk}).

\section{Conclusions}
I hope to have given the impression that while much has been
achieved in classifying supersymmetric solutions of supergravity
theories, I think that much remains to be discovered.

{\bf Acknowledgement} I would like to thank Jan Gutowski, Chris
Hull, Dario Martelli, Stathis Pakis, Harvey Reall, James Sparks
and Dan Waldram for very enjoyable collaborations upon which this
review is based.

\end{document}